\documentclass[]{pasj02} 
\usepackage[switch,mathlines]{lineno} 

\jyear{2025}
\Received{2025 October 14}
\Accepted{2025 December 13}

\usepackage{comment}
\usepackage{natbib}
\usepackage{enumitem}

\newcommand{\na}{\textrm{Nat.\,Astron.}}

\begin{document} 

\title{Bibliometric benchmarking across astronomy journals: Knowledge-use cycle and PASJ in the global landscape}

\author{Hideaki \textsc{Fujiwara}\orcid{0000-0001-6536-8656} }
\affil{Frontier Research Institute for Interdisciplinary Sciences (FRIS), Tohoku University, 6-3 Aramaki-aza Aoba, Aoba-ku, Sendai, Miyagi 980-8578, Japan}
\email{hideaki@fris.tohoku.ac.jp}


\maketitle

\begin{abstract}
We present a comparative bibliometric analysis of eight astronomy journals over 1996--2024, including \textit{Publications of the Astronomical Society of Japan} (PASJ). Using data from Scopus and SciVal, we extract annual indicators of publication activity and scholarly impact, analyze time series, citation distributions, and citation age profiles, and benchmark PASJ within this landscape. 
The age profiles reveal a characteristic knowledge-use cycle: citations rise over $\sim$2--4 years, approach saturation by $\sim$10--12 years, underscoring limits of short-window impact metrics.
Journals published by European and North American astronomical organizations sustain higher impact, whereas PASJ generally lies below the world baseline. 
In parallel, PASJ shows episodic above-baseline impact through facility- or mission-driven special issues and features that, given the journal's modest annual volume, can materially shift year-level metrics. 
These patterns point to two potential avenues for PASJ: well-timed, thoughtfully organized special issues and features that highlight high-impact results, and continued strengthening of international collaboration.
\end{abstract}


\section{Introduction}\label{sec:1}

Scholarly knowledge in astronomy has long been accumulated primarily through peer-reviewed journal articles. Subsequent research builds on prior results, refining, testing, and extending them, and thereby creates a cycle of knowledge use that links earlier publications to new work through citation, synthesis, and replication. This cumulative process motivates a systematic and comparative description of how journals contribute to the dissemination and uptake of astronomical knowledge. Understanding how \textit{Publications of the Astronomical Society of Japan} (\pasj) is positioned relative to other astronomy journals provides insight into the journal's evolving role and impact and informs strategies for enhancing its international visibility. It also offers guidance to researchers seeking greater international reach when publishing in \pasj\ and related venues.

Bibliometrics provides quantitative methods for describing research production, impact, and collaboration across fields and over time. When applied with best-practice guidelines, bibliometric indicators complement qualitative assessment and help contextualize performance at the levels of projects, journals, and countries \citep[e.g.,][]{Moed2010,Bornmann2012,Waltman2016}. For example, standard bibliometric indicators have demonstrated the scholarly impact of publications from the Subaru Telescope's early operational years, with similar assessments conducted for other astronomical facilities as well \citep[and references therein]{Fujiwara2025}.

Within astronomy, early field-specific studies show that citation attention can persist for decades. Longitudinal analyses report a peak around five years after publication, followed by slow decay, yielding long cited half-lives. Observational papers tend to outlast theoretical ones, and influential papers sustain attention longer than controls \citep{Abt1981,Abt1996,Abt2000}. These findings motivate the joint use of short- and long-window indicators when benchmarking journals within a single field.

In this paper, we conduct a comparative bibliometric analysis of peer-reviewed journal articles in astronomy and astrophysics to evaluate and compare output, international collaboration, and scholarly impact for eight journals over 1996--2024. Using publication and citation data from Scopus\footnote{$\langle$https://www.scopus.com/$\rangle$.} and SciVal\footnote{$\langle$http://scival.com/$\rangle$.}, we benchmark \pasj\ within the global landscape by comparing its performance with that of other journals. We also situate journal-level indicators against an area-level baseline for astronomy and astrophysics to contextualize performance.

\section{Data and methodology}\label{sec:2}
\subsection{Dataset construction}\label{ssec:21}

\begin{table*}
  \caption{Target journals, issuing bodies, founding years, and brief scopes.}
  \label{tab:target_journals}
  {\centering
  \small
  \setlength{\tabcolsep}{5pt} 
  \begin{tabular}{p{4.5cm} p{1.1cm} c p{10.0cm}}
    \hline
    \textbf{Journal} & \textbf{Issuing body} & \textbf{Year} & \textbf{Scope} \\
    \hline
    \textit{Astronomy \& Astrophysics} (\aap) & ESO\footnotemark[1] & 1969 & Research across all areas of astronomy and astrophysics. \\
    \textit{The Astronomical Journal} (\aj) & AAS\footnotemark[2] & 1849 & Research emphasizing observational results, surveys, instrumentation, and techniques. \\
    \textit{The Astrophysical Journal} (\apj) & AAS\footnotemark[2] & 1895 & Significant new research, observational, theoretical, or instrumental. \\
    \textit{The Astrophysical Journal Supplement Series} (\apjs) & AAS\footnotemark[2] & 1954 & Extensive papers, catalogs, large surveys/data releases, and software/methods. \\
    \textit{Monthly Notices of the Royal Astronomical Society} (\mnras) & RAS\footnotemark[3] & 1827 & Research in astronomy and astrophysics, observational and theoretical. \\
    \textit{Nature Astronomy} (\na) & Nature Portfolio\footnotemark[4] & 2017 & High-impact research, reviews, and commentary across astronomy, astrophysics, and planetary science. \\
    \textit{Publications of the Astronomical Society of Japan} (\pasj) & ASJ\footnotemark[5] & 1949 & Research across astronomy, astrophysics, and closely related fields. \\
    \textit{Publications of the Astronomical Society of the Pacific} (\pasp) & ASP\footnotemark[6] & 1889 & Research and techniques in astronomy; instrumentation, software, and data analysis. \\
     \hline
  \end{tabular}
  \par}
\begin{tabnote}
\footnotemark[1] The European Southern Observatory. 
\footnotemark[2] The American Astronomical Society. 
\footnotemark[3] The Royal Astronomical Society. 
\footnotemark[4] Published by Springer Nature. 
\footnotemark[5] The Astronomical Society of Japan. 
\footnotemark[6] The Astronomical Society of the Pacific. 
\end{tabnote}
\end{table*}

As briefly reviewed in \citet{Fujiwara2025}, several major bibliographic databases are widely used for literature analysis in astronomy and astrophysics, including the NASA Astrophysics Data System (ADS)\footnote{$\langle$https://ui.adsabs.harvard.edu$\rangle$.}. In this study, we use Scopus as the primary source because it balances coverage across peer-reviewed journals \citep{Mongeon2016}, standardizes subject area and publication type, and underpins widely used benchmarks (e.g., Times Higher Education rankings\footnote{THE Reporters, 2024, World University Rankings 2025: methodology $\langle$https://www.timeshighereducation.com/world-university-rankings/world-university-rankings-2025-methodology$\rangle$.}). These features make Scopus/SciVal suitable for our longitudinal, cross-journal comparisons.

We constructed a journal-level panel dataset from Scopus via SciVal to enable reproducible, cross-journal comparisons under a single harmonized time window. We adopted SciVal's default counting rules (full counting; citations include self-citations) and limited the publication type to peer-reviewed journal articles (Scopus publication type: ``Article''). For a consistent historical span, we fixed the analysis window to 1996--2024, which is the earliest range for which SciVal provides comprehensive field-normalized coverage across the selected titles. All data were retrieved on June 28--30, 2025 from a SciVal snapshot last updated on June 18, 2025.

The target set comprises \pasj\ and seven other astronomy journals indexed in Scopus:
\begin{enumerate}[label=(\arabic*),noitemsep,topsep=0pt,leftmargin=*]
  \item \textit{Astronomy \& Astrophysics} (\aap)
  \item \textit{The Astronomical Journal} (\aj)
  \item \textit{The Astrophysical Journal} (\apj)
  \item \textit{The Astrophysical Journal Supplement Series} (\apjs)
  \item \textit{Monthly Notices of the Royal Astronomical Society} (\mnras)
  \item \textit{Nature Astronomy} (\na)
  \item \textit{Publications of the Astronomical Society of the Pacific} (\pasp)
\end{enumerate}
Table~\ref{tab:target_journals} summarizes the target journals, their issuing bodies, founding years, and brief scopes.
We selected these eight journals because they represent core, field-specific outlets in astronomy and astrophysics with sustained Scopus coverage over 1996--2024, spanning high-volume general journals and influential venue types within the discipline.
\na\ was launched in 2017, so no values are available for 1996--2016 in our time series.

For benchmarking, we also extracted an area-level baseline time series for the Scopus subject category ``Astronomy and Astrophysics'' (All Science Journal Classifications (ASJC) code 3103), hereafter the ``Entire Area.'' The eight journals analyzed are likewise indexed under ASJC 3103, ensuring field consistency between the journal-level series and the area baseline. These choices yield a harmonized year-by-year panel of bibliometric indicators for each journal and support robust longitudinal and cross-journal comparisons of peer-reviewed research output.

\subsection{Bibliometric indicators}\label{ssec:22}

\begin{table*}
  \caption{Bibliometric indicators used in this study.}
  \label{tab:indicators}
  \centering
  \begin{tabular}{lp{11.5cm}}
    \hline
    \textbf{Indicator} & \textbf{Short Description} \\
    \hline
    Scholarly Output & Number of peer-reviewed journal articles of the selected entity. \\
    International Collaboration proportion & Proportion of publications co-authored by institutions from two or more countries. \\
    Citations per Publication (CPP) & Average citation count per publication for the selected entity (Citation Count divided by Scholarly Output). \\
    Field-Weighted Citation Impact (FWCI) & Ratio of citations received to the expected world average for the same ASJC field, publication year, and publication type (world baseline = 1.0). \\
    Top 10\% publication proportion & Proportion of the selected entity's publications that are within the top 10\% worldwide when ranked by FWCI. \\
    \hline
  \end{tabular}
\end{table*}

To assess temporal trends in publication activity and scholarly impact, we extracted annual bibliometric indicators from SciVal for 1996--2024. The indicators are:
\begin{enumerate}[label=(\arabic*),noitemsep,topsep=0pt,leftmargin=*]
  \item Scholarly Output (number of publications)
  \item International Collaboration proportion
  \item Citations per Publication (CPP)
  \item Field-Weighted Citation Impact (FWCI)
  \item Top 10\% publication proportion (SciVal ``Outputs in Top 10\% Citation Percentiles,'' field-normalized)
\end{enumerate}
Each indicator offers a distinct perspective on research performance and is widely used to evaluate scientific productivity, visibility, and influence. Table~\ref{tab:indicators} summarizes these indicators.

For \apj\ and \mnras, which Scopus/SciVal represent as separate ``main'' and ``letter'' editions for part of the period, we aggregate edition-level indicators by a count-weighted average for each publication year. Let $N_m$ and $N_{\ell}$ be the numbers of articles in the ``main'' and ``letter'' editions in that year, and let $I_m$ and $I_{\ell}$ be the corresponding edition-level values of an indicator $I$ (International Collaboration proportion, CPP, FWCI, or the Top 10\% publication proportion). We then compute
\begin{eqnarray}
I_{m+\ell}=\frac{I_m N_m+I_{\ell} N_{\ell}}{N_m+N_{\ell}}.
\end{eqnarray}
This aggregation is exact for the International Collaboration proportion, CPP, and the Top 10\% publication proportion. It is also exact for FWCI in our setting because FWCI is computed per publication year and all articles considered belong to the same ASJC field and publication type, so the expected-citation baseline per paper is common within a year. This procedure yields like-for-like comparisons with journals such as \aap\ and \pasj, where regular and letter-style articles appear under a single integrated title.

\section{Results}\label{sec:3}
\subsection{Scholarly output}\label{ssec:31}

\begin{figure*}
 \begin{center}
  \includegraphics[width=150mm]{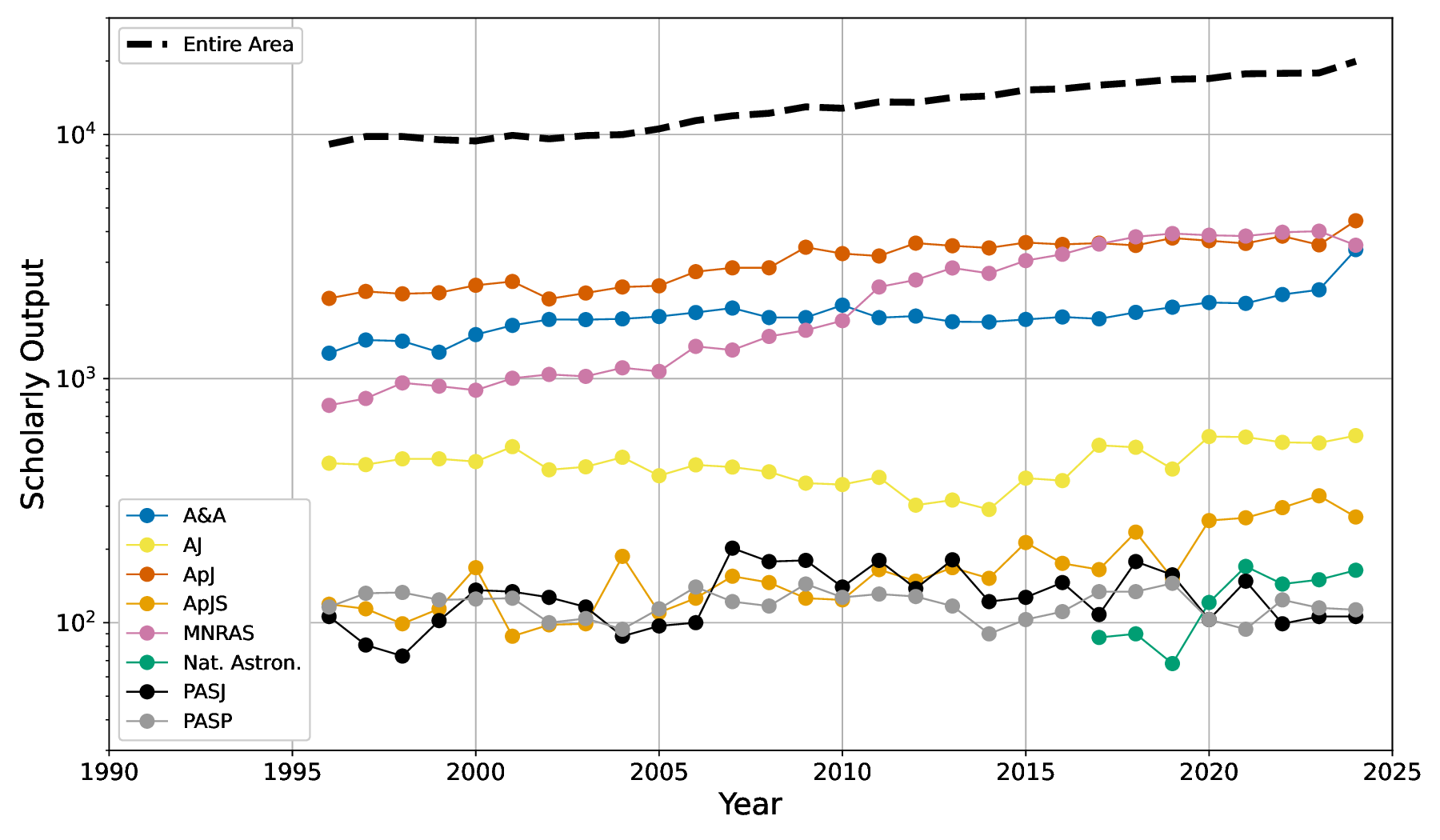}
 \end{center}
   \caption{Annual number of peer-reviewed articles (Scholarly Output) published from 1996 to 2024 in selected astronomy journals. The black dashed line represents the total scholarly output in the ``Entire Area'' as indexed in Scopus. The vertical axis is logarithmic.
{Alt text: A line graph with a logarithmic vertical axis. The x-axis shows publication years from 1996 to 2024. The y-axis represents the number of peer-reviewed journal articles, ranging from about 30 to about 30,000. Eight colored lines trace individual journals, and a black dashed line shows the ``Entire Area'' total.}   
}
  \label{fig:output}
\end{figure*}

Figure~\ref{fig:output} shows annual scholarly output for eight journals and the ``Entire Area'' baseline. 
The series are limited to peer-reviewed articles in SciVal and capture broad patterns in scholarly productivity over nearly three decades.
The total number of articles in the ``Entire Area'' remained relatively stable at around 10,000 per year until 2005, after which a steady upward trend began. By the early 2020s, the annual publication volume reached approximately 17,000--18,000 articles, representing a growth factor of about 1.7--1.8 compared to the late 1990s.

Among the eight journals analyzed, \apj, \mnras, and \aap\ have consistently recorded the highest annual scholarly output. In particular, \mnras\ expanded from $\sim$900 articles per year in the late 1990s to $\sim$4,000 in the early 2020s, with a marked increase between 2010 and 2011 (from 1,727 to 2,373). \aj\ and \pasp\ maintained relatively stable annual volumes throughout the period, at approximately 300--600 and 100--150 articles per year, respectively. \apjs\ shows a lower overall output but exhibits substantial year-to-year fluctuations, including notable peaks. \na\ has among the lowest annual volumes, averaging about 100--170 articles in recent years.

\pasj\ published approximately 100--180 articles per year across most of the study period. Unlike journals that mirrored the overall growth observed in the ``Entire Area,'' \pasj\ does not exhibit a clear long-term upward trend in annual output. However, its publication volume temporarily approached $\sim$180 articles (roughly 1.8 times its late-1990s level) during 2007--2013 and again in 2018--2019.

\subsection{International collaboration proportion}\label{ssec:32}

\begin{figure*}
 \begin{center}
  \includegraphics[width=150mm]{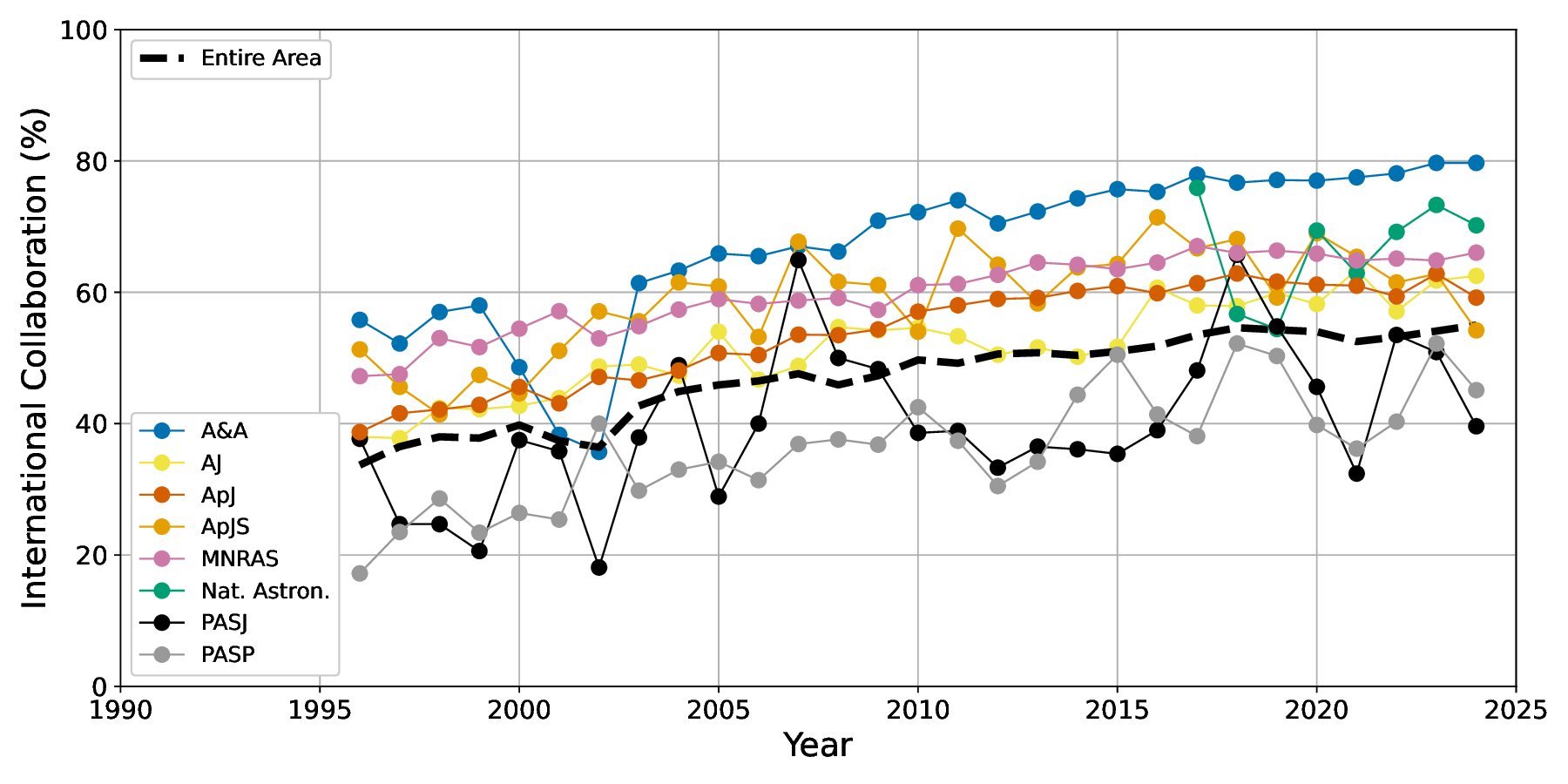}
 \end{center}
\caption{Annual share of internationally co-authored peer-reviewed articles (International Collaboration) from 1996 to 2024 in selected astronomy journals. \aap\ shows the highest and most sustained collaboration rates, followed by \mnras, \apj, and \aj; \na\ has remained high since 2017. \pasj\ records the lowest values across the period, and \pasp\ tracks slightly higher levels. The black dashed line marks the ``Entire Area'' average. The vertical axis is linear and reports percentages.
{Alt text: A line graph with a linear vertical axis. The x-axis shows publication years from 1996 to 2024. The y-axis shows the share of internationally co-authored papers (percent), ranging from 0\% to 100\%. Multiple colored lines represent individual journals and a black dashed line shows the ``Entire Area'' average. }   
}  \label{fig:intl}
\end{figure*}

Figure~\ref{fig:intl} shows the annual proportion of internationally co-authored articles (two or more countries).
Among the journals examined, \aap\ shows the highest and most sustained levels of international collaboration. Since the early 2000s, more than 70\% of its articles have involved cross-country co-authorship, reaching around 80\% in recent years. \mnras, \apj, and \aj\ also display clear upward trends over the study period, rising from roughly 30--40\% in the late 1990s to around or above 60\% by the early 2020s. \na, launched in 2017, immediately showed high collaboration rates around 60--70\% and has maintained similar levels thereafter. \apjs\ exhibits larger year-to-year variation and no strong long-term trend, but generally falls in the 40--60\% range.

\pasj\ has the lowest international collaboration among the journals considered. Although its share has risen from below 20\% in the early 2000s to roughly 30--35\% in recent years, it remains below both the field-wide average and the levels observed for other journals. Because \pasj\ publishes fewer articles per year than high-volume titles, it shows larger year-to-year fluctuations, including occasional spikes.

\pasp\ follows a trajectory similar to \pasj, typically ranging between 30\% and 45\% across the period. The field-wide average (the ``Entire Area'' line) also increases gradually, from about 35\% in the late 1990s to just above 50\% in recent years. Overall, the results indicate substantial differences in the degree of internationalization across astronomy journals.

\subsection{Citations per publication}\label{ssec:33}

\begin{figure*}
 \begin{center}
  \includegraphics[width=150mm]{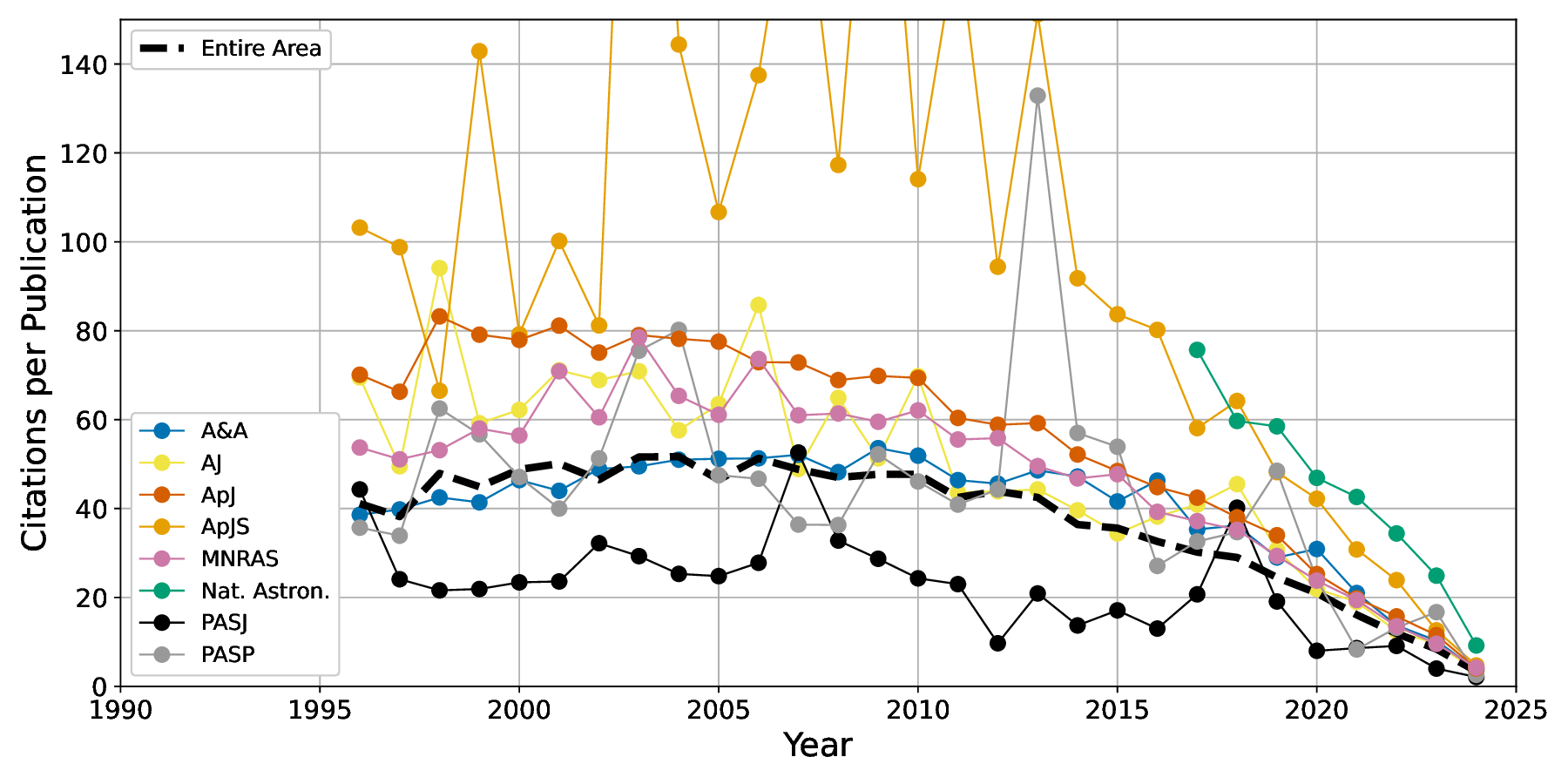}
 \end{center}
   \caption{Annual trends in CPP from 1996 to 2024 across major astronomy journals. CPP values remain relatively stable for articles published before 2010, but decline sharply for more recent publication years. This pattern reflects the effects of citation aging and temporal bias: articles typically accumulate citations over a 10--15 year window, and recent papers have not yet had time to reach their full citation potential. These results highlight the importance of accounting for time-based citation dynamics when interpreting bibliometric indicators.
{Alt text: A line chart showing Citations per Publication (CPP) by publication year from 1996 to 2024 for eight astronomy journals and an Entire Area baseline. }
}
  \label{fig:cpp}
\end{figure*}

We now present results on citation impact indicators. We begin with CPP as a baseline that provides an intuitive, non-normalized view of citation volumes. Because all titles analyzed belong to the same ASJC in Scopus, cross-field differences are minimized. Within a fixed publication year, CPP therefore serves as a reasonable first-order proxy for relative citation impact. Formal comparative assessments are based on field-normalized indicators (FWCI and the Top 10\% publication proportion) in the following subsections.

Figure~\ref{fig:cpp} shows annual CPP trends for the eight journals from 1996 to 2024. At the field level, the ``Entire Area'' baseline peaks for publication years in the mid-2000s and then declines for more recent years, reflecting the limited time available for citations to accrue.

Among individual titles, \apj\ typically attains higher CPP for older publication years, often comparable to or slightly above \mnras\ and \aap. \mnras\ shows sustained CPP through the late 2000s, followed by a gradual softening for later publication years; \aap\ follows a similar profile at a slightly lower level. \aj\ generally occupies an intermediate range between the largest general journals and lower-volume society journals. \apjs\ exhibits the largest year-to-year variability, with pronounced peaks in some years. \pasj\ and \pasp\ tend to lie in the lower CPP range for older publication years; \pasp\ also shows occasional elevated years. \na\ shows comparatively high CPP for its earliest publication year, followed by lower values for newer years that have had less time to accumulate citations.

\subsection{Field-weighted citation impact}\label{ssec:34}

\begin{figure*}
 \begin{center}
  \includegraphics[width=150mm]{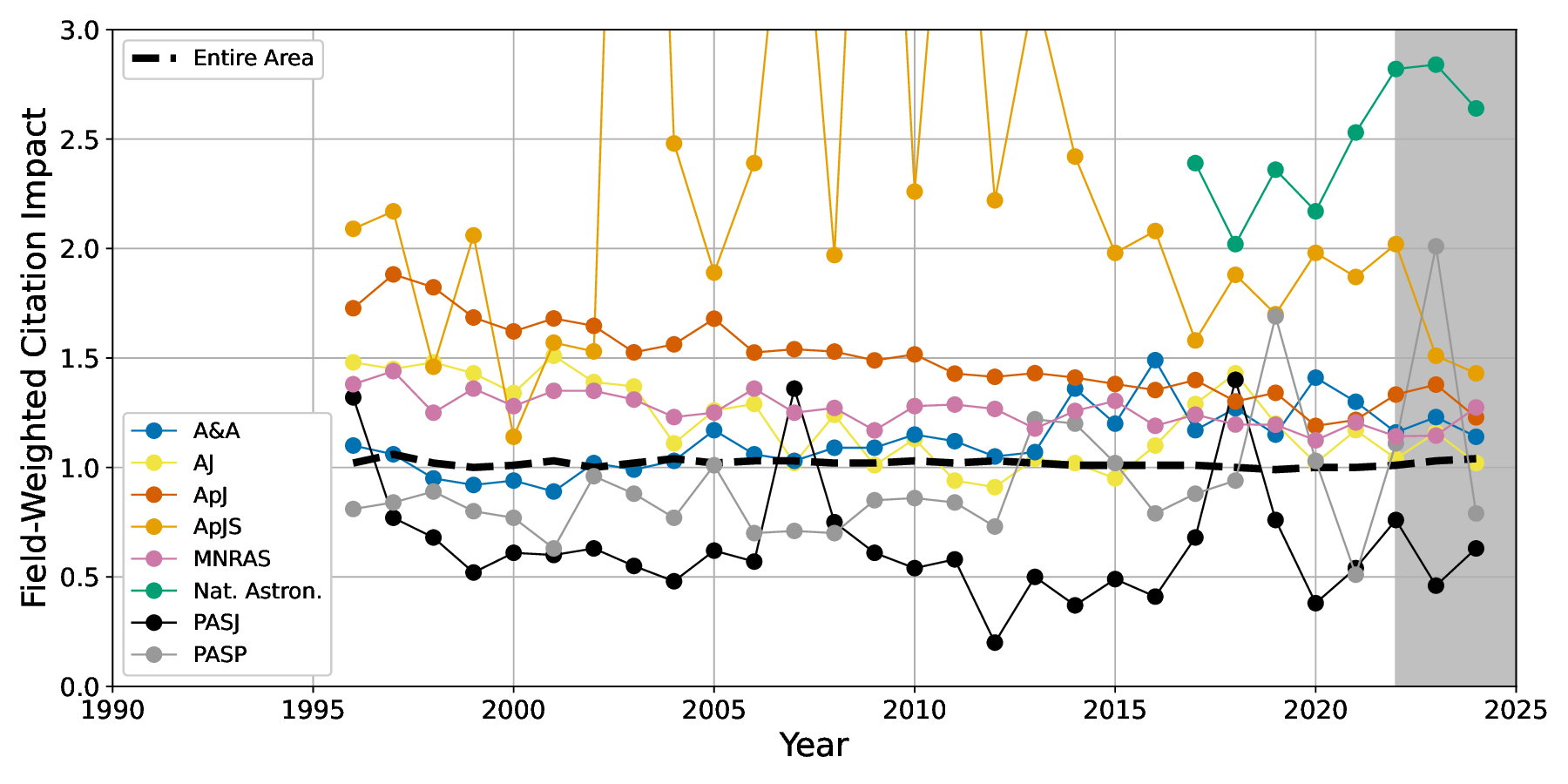}
 \end{center}
\caption{
Annual FWCI for selected astronomy journals from 1996 to 2024. FWCI is computed from citations received during the publication year and the subsequent three calendar years and is normalized by field, publication year, and publication type; 1.0 corresponds to the world average. Values for 2022 and later (shaded in gray) are provisional as of June 2025 because the four-year window is incomplete.
\na\ shows the highest and most consistent FWCI values since its launch. \apjs\ exhibits large annual variation but frequently exceeds 2.0. In contrast, \pasj\ remains consistently below the global average. 
{Alt text: A line graph with a linear vertical axis showing FWCI trends for eight astronomy journals from 1996 to 2024. Each line represents a journal. Values for 2022 onward are shaded in gray to indicate that citation windows are incomplete. \na\ shows the highest values, followed by \apjs. \pasj\ remains consistently below the average.}
}
  \label{fig:fwci}
\end{figure*}

Figure~\ref{fig:fwci} shows annual FWCI with the ``Entire Area'' near 1.0 by definition. Note that FWCI uses citations in the publication year plus three subsequent years, and therefore, values from 2022 onward are based on incomplete windows.

Among the journals analyzed, \apjs\ consistently shows the highest FWCI values, often exceeding 2.0, though its year-to-year variation is substantial, including an extreme peak of $\sim 7.2$ in 2003. \na, since its launch in 2017, has maintained both high and stable FWCI levels, ranging from 2.0 to 2.8. \apj\ stood clearly above the others (except \apjs) in the late 1990s but has shown a gradual downward trend over the decades, declining from about 1.7--1.8 to around 1.2 in 2020--2023. In contrast, \aap\ has exhibited a noticeable increase since around 2014, reaching values in the range of 1.2--1.5. \mnras\ and \aj\ have shown stable FWCI trends, typically fluctuating between 1.0 and 1.4. \pasp\ generally remains below 1.0 but displays notable spikes in 2019 and 2023.

In comparison, \pasj\ has consistently recorded FWCI values below the global average. Throughout most of the period, it ranges between 0.4 and 0.8. Two exceptional years, 2007 and 2018 (both $\sim 1.4$), saw temporary surges above 1.0, but no sustained upward trend is observed.

\subsection{Top 10\% publication proportion}\label{ssec:35}

\begin{figure*}
 \begin{center}
  \includegraphics[width=150mm]{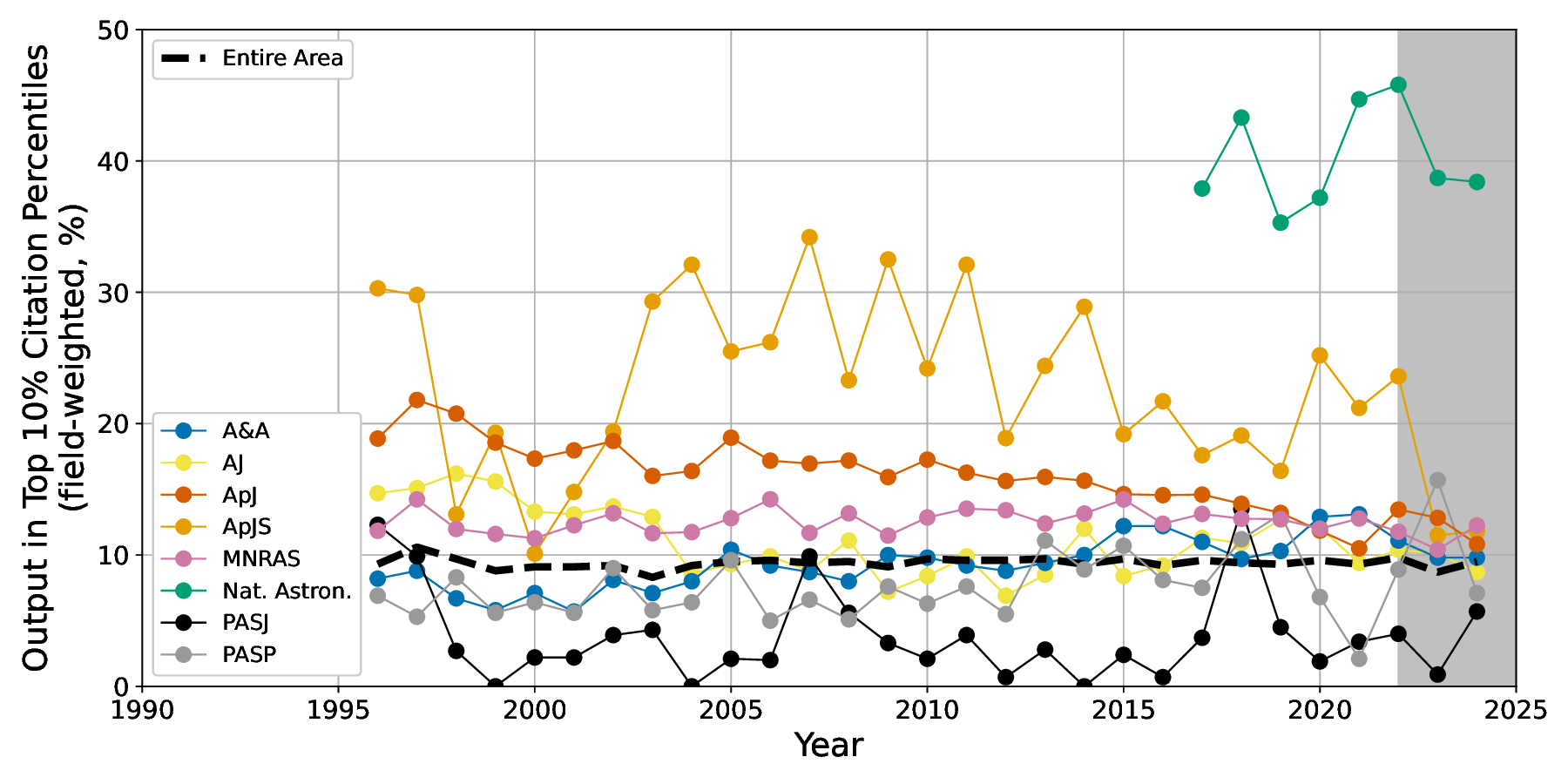}
 \end{center}
  \caption{Annual Top 10\% publication proportion for each journal, 1996--2024. \apj\ and \mnras\ consistently maintain higher proportions, typically between 30--40\%, while \pasj\ and \pasp\  remain below the global average throughout the period. These results reflect differences in overall citation impact and visibility across journals.
  {Alt text: A line graph with a linear vertical axis showing trends of annual Top 10\% publication proportion for eight astronomy journals from 1996 to 2024. Each line represents a journal. Values for 2022 onward are shaded in gray to indicate that citation windows are incomplete. \na\ shows the highest values, followed by \apjs. \pasj\ remains consistently below the average.}
}
  \label{fig:top10}
\end{figure*}

Figure~\ref{fig:top10} reports each journal's Top 10\% publication proportion in the global astronomy and astrophysics field. This metric complements FWCI by indicating how frequently a journal publishes highly cited papers.

Among the journals analyzed, \mnras\ and \apj\ maintain consistently high Top 10\% publication proportions across the study period, typically ranging between 30\% and 40\%. \aap\ follows closely, with similarly strong performance in most years. \na\ stands out with exceptionally high values, regularly exceeding 50\% despite its relatively short publication history, consistent with its high FWCI.
\apjs\ exhibits more variable behavior, with several sharp peaks corresponding to specific years. These peaks are often associated with the release of high-profile data sets or mission-related papers, such as those from the Wilkinson Microwave Anisotropy Probe \citep[e.g.,][]{Spergel2003} and the Sloan Digital Sky Survey \citep[e.g.,][]{Abazajian2009}.

By contrast, \pasj\ and \pasp\ generally remain below the field baseline (10\%) for the Top 10\% publication proportion. For \pasj, the share of top-cited papers rarely exceeds 10\% and often falls below 5\%. This pattern is consistent with a lower degree of internationalization and with more limited participation in large collaborations (see Section~\ref{ssec:32}).

These results align with the broader patterns seen in FWCI and other citation indicators: journals such as \mnras, \apj, and \na\ regularly contribute to the most-cited segment of the literature, whereas \pasj\ occupies a more modest citation standing in the global landscape.

\subsection{Citation distribution by journal and period}\label{ssec:36}

\begin{figure*}
 \begin{center}
  \includegraphics[width=85mm]{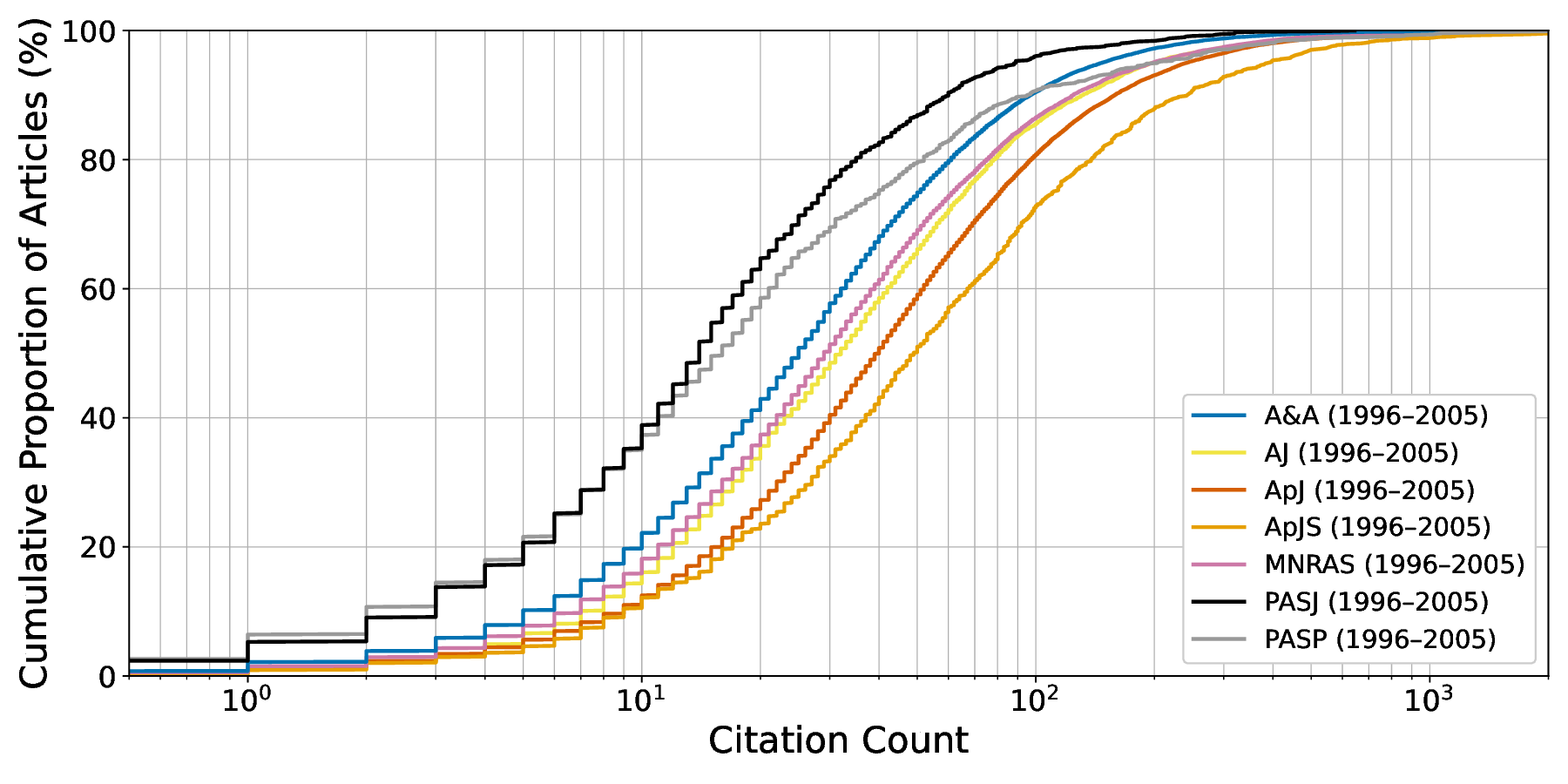}
  \includegraphics[width=85mm]{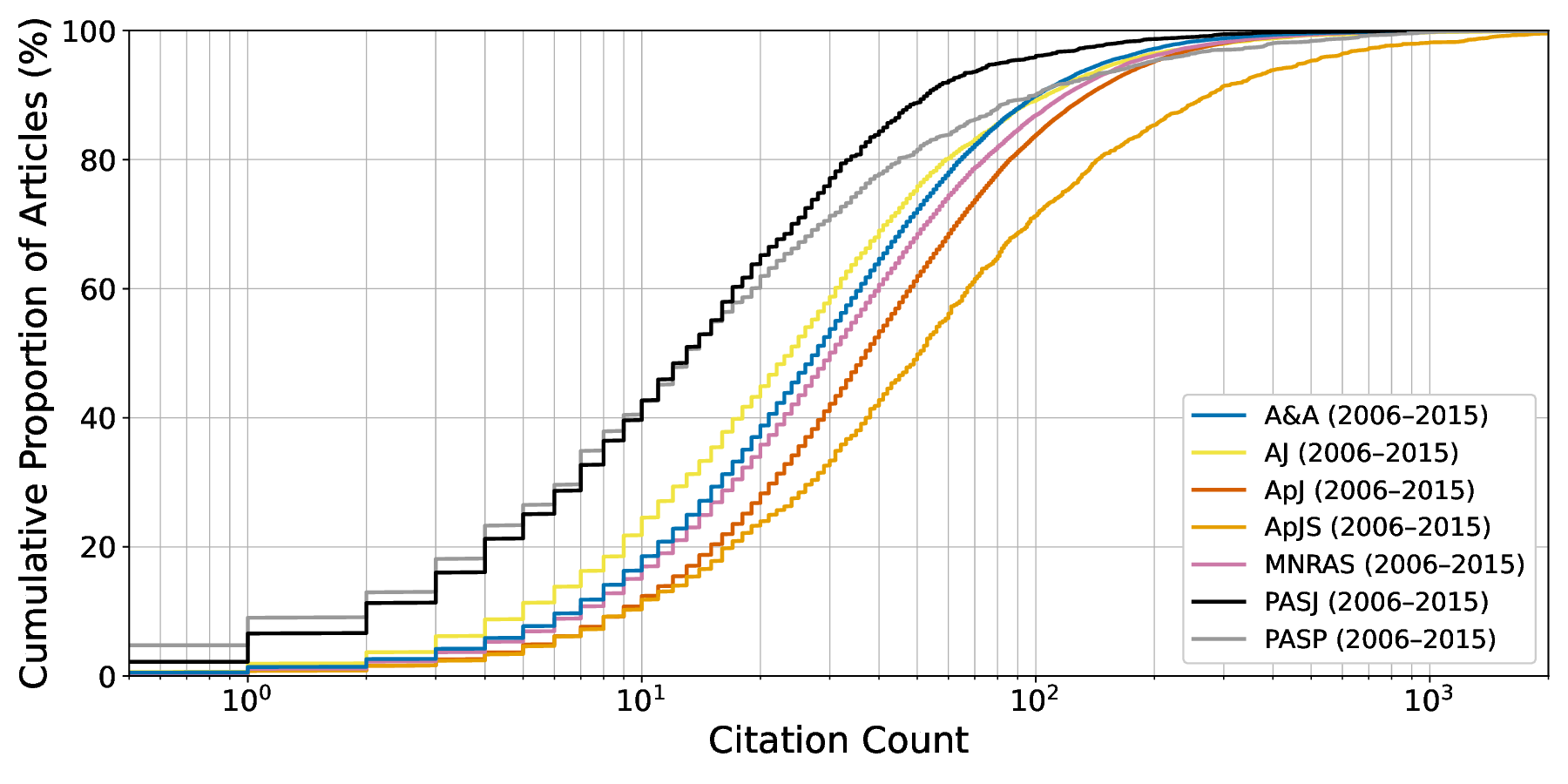}
  \includegraphics[width=85mm]{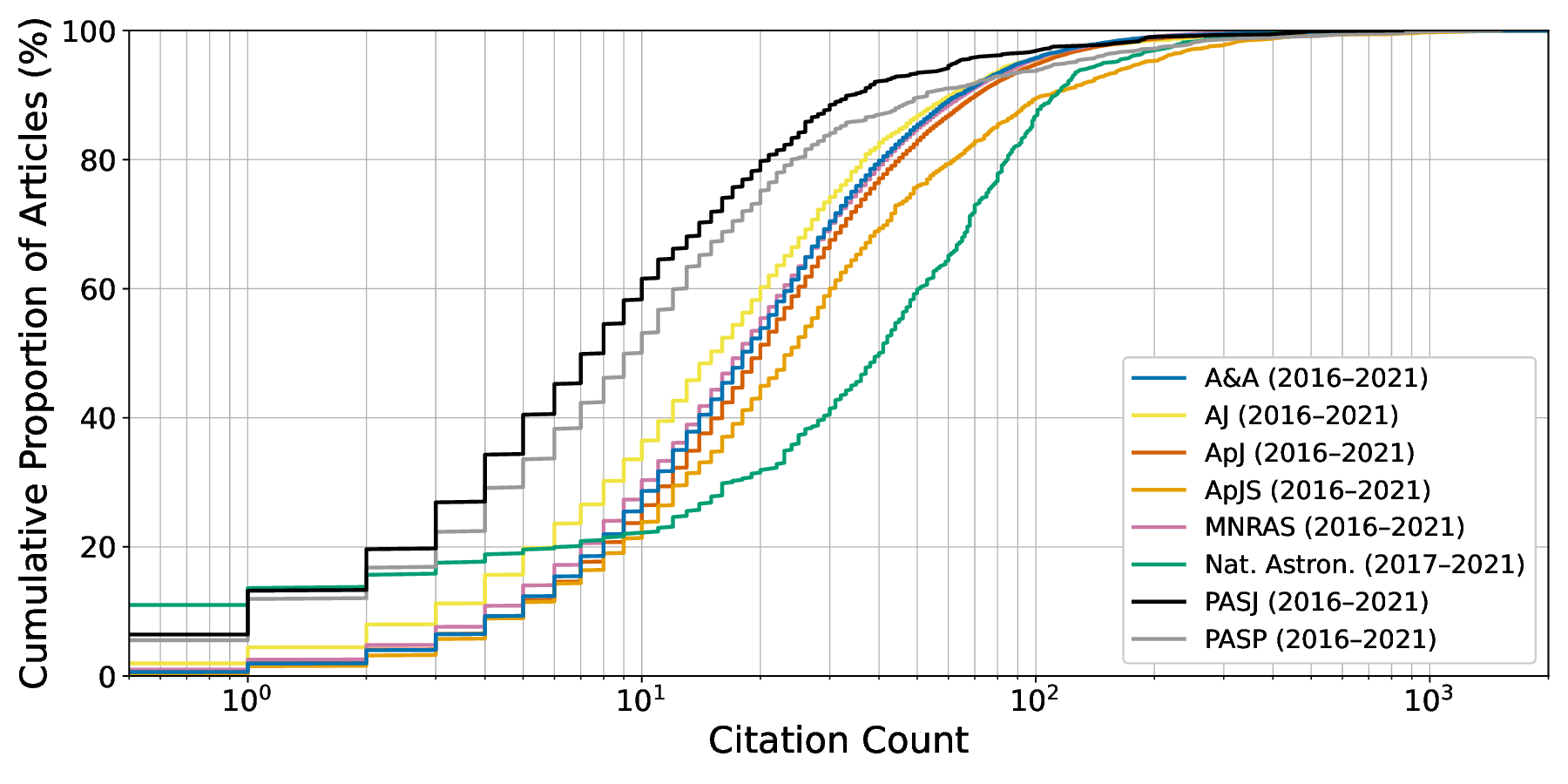}
  \includegraphics[width=85mm]{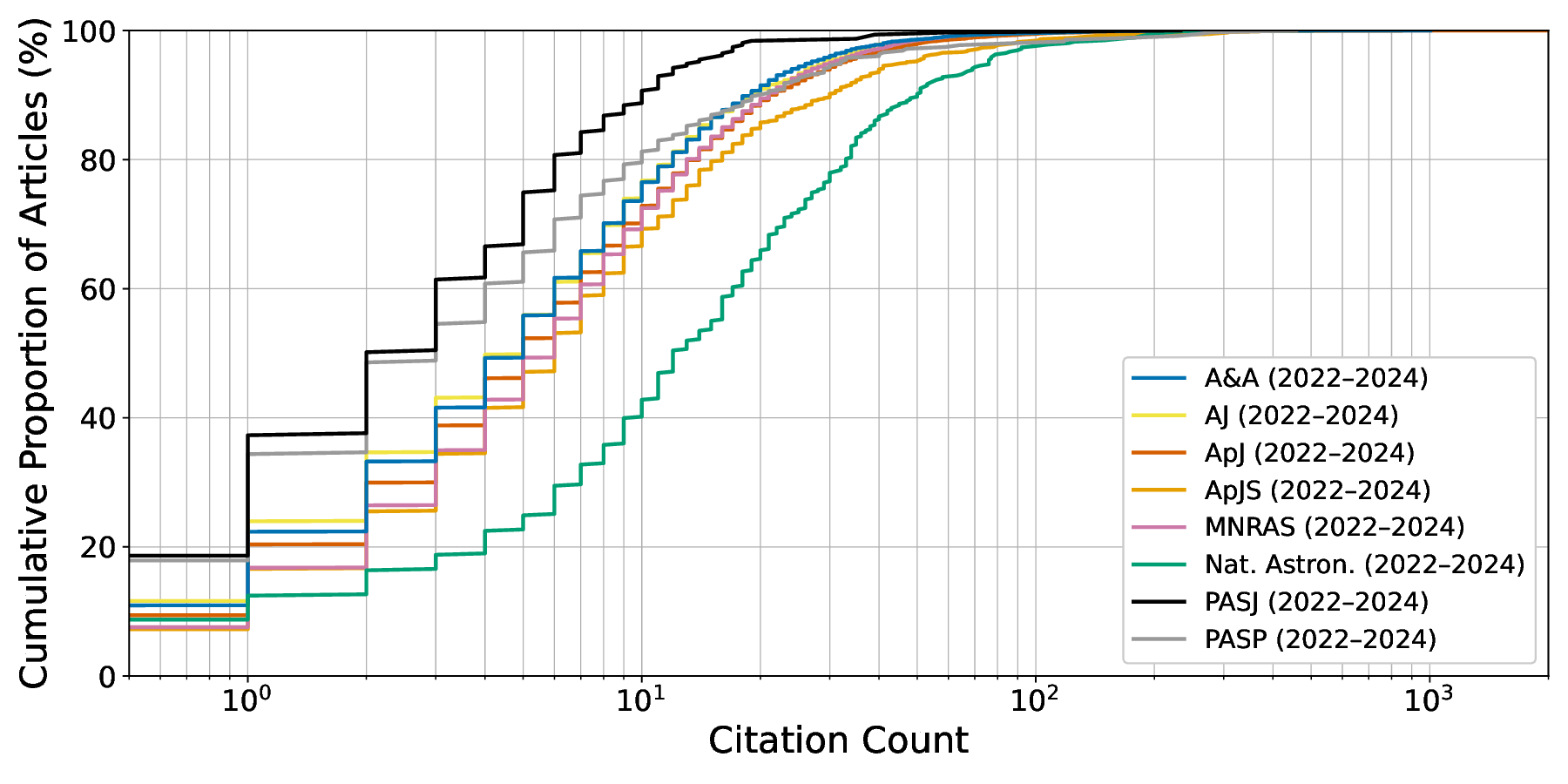}
 \end{center}
  \caption{Cumulative citation distributions of peer-reviewed research articles by journal for four publication periods (1996--2005, 2006--2015, 2016--2021, and 2022--2024). The vertical axis shows the cumulative proportion of articles and the horizontal axis shows citation count on a logarithmic scale. On the logarithmic x-axis, values plotted to the left of $10^{0}$ represent zero-citation articles. Curves further to the right indicate a larger share of highly cited articles. \na\ appears only in panels from 2016 onward. Distributions for 2022--2024 are provisional because citation windows are still incomplete.
{Alt text: Four panel line charts showing cumulative percentages of articles versus citation counts on a log scale for major astronomy journals. In earlier periods, \apj, \mnras, and \aap\ are right-shifted relative to \pasj\ and \pasp; the most recent panel shows steeper rises at low counts for all journals.}
}
  \label{fig:citation_cumulative}
\end{figure*}

To complement average-based indicators, we examine cumulative citation distributions across four publication windows (Figure~\ref{fig:citation_cumulative}). This view exposes the full spread and upper tail of citations while keeping exposure roughly comparable within each window. Curves farther to the right indicate a larger share of highly cited papers, whereas a steep rise near low counts indicates many low-citation articles.

For 1996--2015, when citation windows are largely mature, \apj, \mnras, and \aap\ are right-shifted with long upper tails, while \pasp\ lies between these journals and \pasj. The \pasj\ curve rises steeply at low counts and reaches unity earlier, implying a smaller upper-tail mass. The ordering in 2006--2015 is similar: \apj\ and \mnras\ retain broad distributions with extended right tails, \aap\ is close but slightly left-shifted, and \pasp\ remains intermediate. \pasj\ shifts modestly to the right relative to 1996--2005 but still trails the largest general journals.

In 2016--2021, \na\ appears and is relatively right-shifted within this shorter-accumulation window. \apj, \mnras, and \aap\ continue to show broader distributions than \pasj\ and \pasp, and \pasj\ remains the leftmost curve. In 2022--2024, all curves compress toward low counts because citations are still accruing, shortening the upper tails for every journal. Even so, relative positions largely persist, with \apj, \mnras, and \aap\ to the right of \pasp\ and \pasj.

Across all periods, \pasj\ consistently shows (i) a steeper initial rise at low citation counts, (ii) earlier convergence toward unity, and (iii) a thinner upper tail than \apj, \mnras, and \aap. These features indicate lower median and upper-percentile citation levels for \pasj\ than for the largest general journals. A modest rightward movement is visible from 1996--2005 to 2006--2015, but the cross-journal ordering is otherwise stable.

\section{Discussion}\label{sec:4}
\subsection{Key findings in context}\label{ssec:41}

This study compared eight astronomy journals in 1996--2024 using output, collaboration, and citation indicators within a single subject field. We identify three principal patterns.
First, publication volumes grew field-wide, but journal trajectories diverged. \apj, \mnras, and \aap\ had the largest annual outputs. \pasj\ stayed around 100--180 research articles per year without a sustained upward trend.

Second, international co-authorship rose across the field. \aap\ reached about 70--80\% in recent years. \mnras, \apj, and \aj\ increased from roughly 30--40\% in the late 1990s to around or above 60\% by the early 2020s. \pasj\ remained lower, rising from below 20\% to about 30--35\%, and \pasp\ followed a similar but slightly higher path. These figures indicate heterogeneous internationalization.

Third, citation indicators show consistent ordering. In CPP for older publication years, \apj, \mnras, and \aap\ lie above \pasj\ and \pasp, while \apjs\ varies strongly year to year. In FWCI, \na\ has been high and stable since 2017; \apjs\ is often $>2.0$ but volatile; \apj\ declines from about 1.7--1.8 to $\sim$1.2 in 2020--2023; \aap\ rises to roughly 1.2--1.5; and \pasj\ is usually below 1.0. The volatility of \apjs\ likely reflects a small annual denominator and the outsized impact of a few data-release or catalog-style papers, consistent with heavy-tailed citation distributions. The Top 10\% publication proportion shows the same ordering, with \pasj\ typically well below the field average.

Distributional evidence reinforces these patterns. Cumulative citation curves for 1996--2005 and 2006--2015 are right-shifted with heavier tails for \apj, \mnras, and \aap\ relative to \pasj\ and \pasp. In 2016--2021, \na\ is comparatively right-shifted. In 2022--2024, all curves compress toward low counts because citations are still accruing, yet the relative ordering persists. Together, these results indicate persistent upper-tail differences across journals, consistent with the age profiles in Section~\ref{ssec:42}.

\subsection{Interpreting citation metrics over time}\label{ssec:42}

\begin{figure*}
 \begin{center}
  \includegraphics[width=150mm]{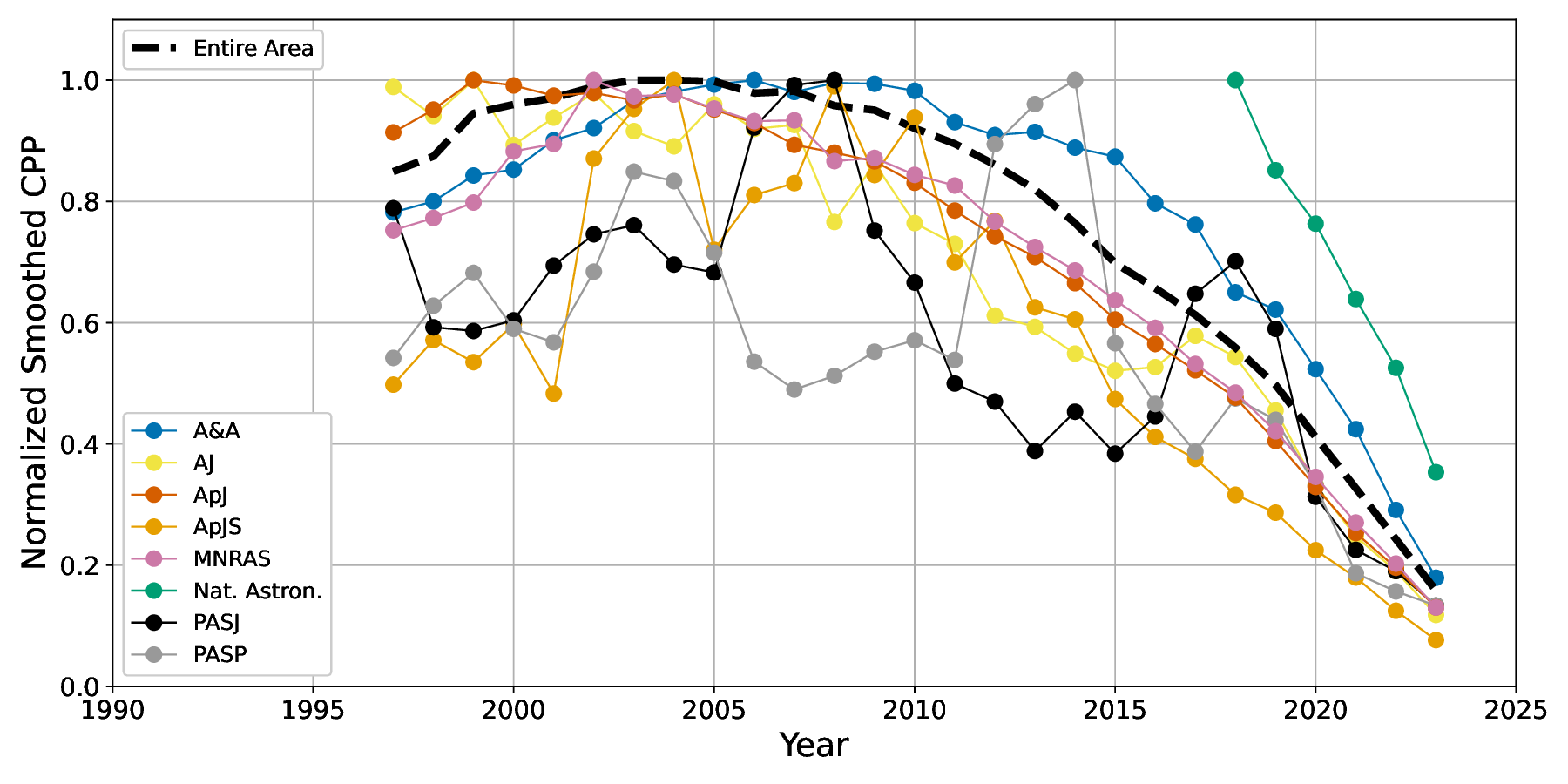}
 \end{center}
   \caption{Normalized citations per publication (CPP) by calendar year, 1996--2024, for eight major astronomy journals. For each journal, annual CPP values are smoothed with a 3-year moving average and then divided by that journal's own peak CPP (1996--2024), emphasizing trajectory shape independent of absolute scale. The moving average is centered in time; edge years use a shorter window. Normalization facilitates comparing the rise and persistence of attention across titles. Values in the most recent years should be interpreted with caution because citations are still accruing and moving-average edge effects are larger.
{Alt text: A line graph. The x-axis shows years from 1996 to 2024. The y-axis shows normalized citations per paper (0--1), using a 3-year moving average. Eight colored lines represent individual journals.}
}
  \label{fig:cpp_smoothed_normalized}
\end{figure*}

\begin{figure*}
 \begin{center}
  \includegraphics[width=150mm]{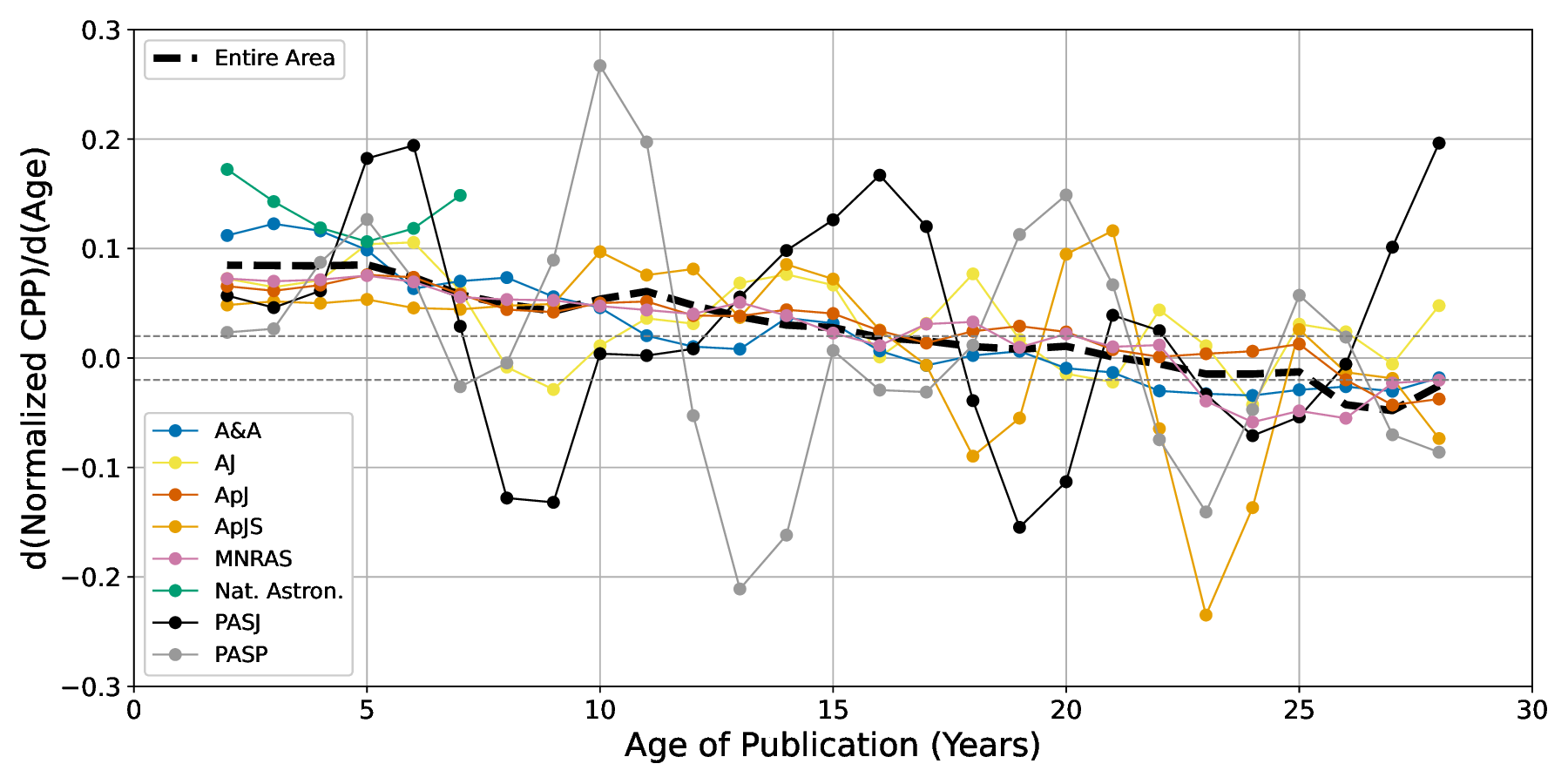}
 \end{center}
   \caption{Age profile of citation growth for eight major astronomy journals (ages measured as of 2025; i.e., $\text{age} = 2025 - \text{publication year}$). For each journal, the annual change of smoothed and normalized CPP, or $d$(normalized CPP)/$d$(age), is plotted. Positive values indicate continued growth, values near zero indicate saturation, and negative values indicate decline. Growth typically peaks within $\sim$2--4 years and approaches $\sim$0 by $\sim$10--12 years; later ages are increasingly censored by limited observation windows. Interpret the youngest and oldest ages with caution due to edge effects and incomplete citation accrual. 
{Alt text: A line graph. The x-axis shows age since publication (years). The y-axis shows the annual change in normalized citations per paper. Eight colored lines represent individual journals.}
}
  \label{fig:cpp_normalized_gradient_by_age_all_journals}
\end{figure*}

In light of the time-series patterns in Section~\ref{sec:3} and the short-window caveats in Section~\ref{ssec:41}, our results show that citation impact evolves on multiyear timescales. To separate ``shape'' from ``level,'' Figure~\ref{fig:cpp_smoothed_normalized} plots each journal's CPP normalized by its own peak and smoothed with a centered three-year moving average to reduce year-to-year noise. This facilitates comparison of how quickly attention rises and how long it persists, independent of absolute volumes. As a complementary view, Figure~\ref{fig:cpp_normalized_gradient_by_age_all_journals} re-expresses the horizontal axis as age since publication and, after the same smoothing, displays the annual change as a function of age. Positive values indicate continued growth, values near zero indicate saturation, and negative values indicate decline.

The two figures indicate a broadly similar temporal pattern across journals. Normalized CPP typically increases for the first several years and then flattens. In the age-based plot, growth is strongest within about 2--4 years, declines steadily thereafter, and approaches zero by roughly 10--12 years for most titles. Residual growth beyond year 12 is usually small. These timings are consistent with a long but finite attention horizon. Note also that the figures summarize journals at the aggregate level, and the youngest ages are right-censored by construction; individual articles can deviate substantially from the average path.

Differences among journals are mostly in amplitude rather than timing. \apj, \mnras, and \aap\ reach higher normalized levels and maintain them over longer spans, whereas \pasj\ and \pasp\ peak at lower levels but approach saturation on similar timescales. \apjs\ shows sharper local increases at some ages, reflecting episodic surges associated with specific article types. \na\ is still young within our window, so its age profile has not yet fully stabilized.

These dynamics help interpret the other indicators in this paper. FWCI uses citations in the publication year and the three subsequent calendar years, which captures the early rise but not the extended accumulation that continues for many astronomy papers. CPP measured within a fixed publication year provides an intuitive baseline for relative impact among journals in the same field, but recent years are systematically low because articles have had less time to gather citations. Accordingly, short-window indicators should be treated as provisional for the most recent years (e.g., 2022--2024 in our data).

A practical implication is that astronomy's knowledge-use cycle is long. A working summary from our data is an early attention phase in the first 3--4 years, a diffusion and consolidation phase to about 10--12 years, and a slow tail thereafter up to roughly 15 years. Evaluations that rely only on very short windows can understate eventual influence, while windows of about a decade align better with typical maturation of citation impact. The observed maturation window may relate to the field's multiyear research cadence, including major observing programs, mission operations, and facility commissioning cycles. A rigorous examination of these linkages represents a promising avenue for future work.

The notion of a citation ``half-life'' was introduced to characterize literature obsolescence \citep{BurtonKebler1960} and is formalized in Journal Citation Reports\footnote{$\langle$https://jcr.clarivate.com/$\rangle$.} as cited and citing half-life metrics. Within astronomy, early field-specific work already noted long citation half-lives. \citet{Peterson1988} documented extended citation longevity in astronomical journals and cautioned that short citation windows can understate eventual impact. Our age profiles are consistent with this picture and reinforce the value of estimating journal- and subfield-specific half-lives within a harmonized sampling frame.

Additional evidence within astronomy comes from longitudinal analyses of citation histories. \citet{Abt1981} examined papers published in 1961 and found that citations typically peak about five years after publication and then decline slowly, reaching roughly half the peak by year twenty. \citet{Abt1996} tracked papers over four decades and reported long cited half-lives, with observational papers outlasting theoretical ones. \citet{Abt2000} further showed that `important' papers maintain citation attention far longer than controls. Our age profiles align with these patterns and underscore the need to pair short-window metrics with indicators that reflect extended trajectories.

While comparisons across disciplines beyond astronomy and astrophysics are of clear interest, they require harmonized sampling frames beyond the present study. We therefore confine our interpretation of the knowledge-use cycle to astronomy and astrophysics and leave systematic cross-field comparisons to future work.

\subsection{FWCI--CPP isochrones: evidence of short-window bias}\label{ssec:43}

\begin{figure*}
\begin{center}
\includegraphics[width=130mm]{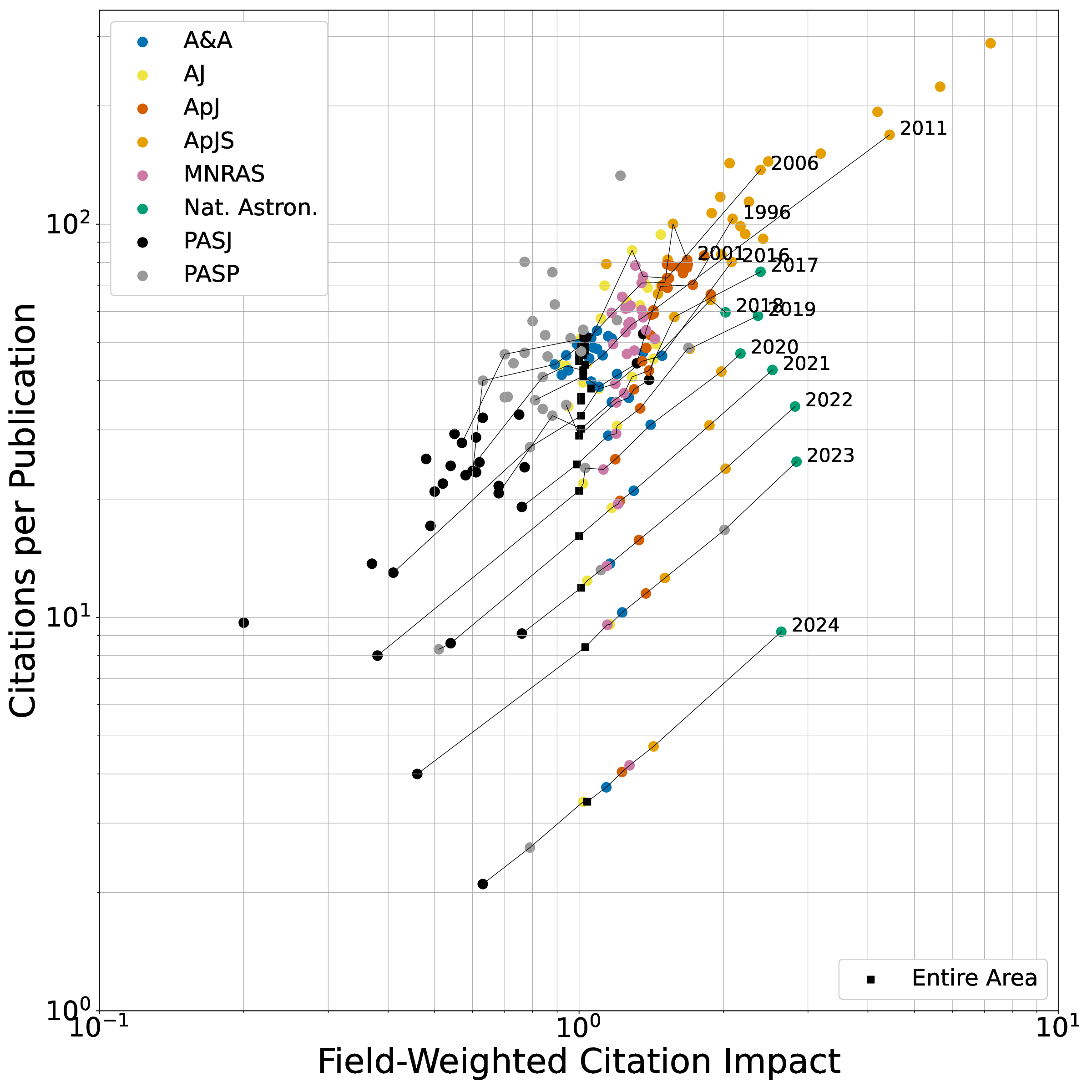}
\end{center}
\caption{FWCI versus CPP for eight astronomy journals, 1996--2024. Each point is a journal's annual metrics; year-wise isochrones (thin lines) connect journals from the same publication year. Both axes are logarithmic. Isochrones from recent years are nearly linear, consistent with the uniform short-window nature of FWCI. Older isochrones are more irregular, reflecting the divergence between short-window impact and cumulative citation accrual. Black squares mark the field baseline (``Entire Area'').
{Alt text: A log-log scatterplot of FWCI (x) versus CPP (y) by journal and year, with thin lines connecting journals from the same year; recent-year lines are nearly straight, older-year lines are more irregular.}
}
\label{fig:fwci-cpp_isochrone}
\end{figure*}

To complement the time-series analysis in Section~\ref{ssec:42}, Figure~\ref{fig:fwci-cpp_isochrone} shows a two-dimensional view of citation performance by plotting FWCI against CPP for eight astronomy journals. Each marker is a journal--year observation. Year-specific ``isochrones'' (equal-publication-year polylines) connect points from the same publication year, enabling cross-sectional comparisons across time.

For recent years (e.g., 2020--2024), isochrones are nearly linear and tightly aligned. This coherence is expected because FWCI evaluates citation performance within a short window (publication year plus the subsequent three calendar years) and is normalized by field, year, and publication type. In contrast, older isochrones (e.g., 1996--2010) appear increasingly irregular. As citations continue to accrue over longer horizons, CPP grows, while FWCI, by design, reflects citations accrued only within that short window for each publication year and therefore changes little relative to CPP. The growing dispersion in older isochrones highlights the same age-dependent dynamics noted in Section~\ref{ssec:42}: early growth over $\sim$2--4 years, near-saturation by $\sim$10--12 years, and a slow tail thereafter.

This pattern implies that relying on short-window metrics alone can understate long-term influence, particularly for venues with steady, durable citation accrual. Indicators that reflect extended trajectories (such as CPP) are therefore essential complements when assessing journals across many publication years. This consideration extends to article-level evaluation: short-window metrics such as FWCI may systematically understate the long-term influence of papers whose citations accrue over a decade or more; a systematic article-level analysis is left for future work.

\subsection{PASJ's current position and the episodic amplification effect of special issues and features}\label{ssec:44}

\begin{table*}
\caption{List of Special Issues and Features in PASJ (1996--2023) with average FWCI of $\gtrsim 1.0$.}
\label{tab:pasj_special_issues}
{\centering
\begin{tabular}{llcccp{11.5cm}}
\hline
Year & Vol.,  & \# of     & \# of          & FWCI  & Theme of Special Issues/Features \\
        & Issue & Papers\footnotemark[1] & Articles\footnotemark[2]           &            & \\
\hline
1996 & 48, 2--3 & 19 & 19 & 3.14 & Recent Results from ASCA \\
1996 & 48, 5--6 & 9 & 9 & 1.03 & Recent Results from IRTS \\
2002 & 54, 6 & 14 & 14 & 1.52 & Recent Results from the Subaru Telescope \\
2005 & 57, 1 & 9 & 9 & 0.97 & Recent Progress of Stellar Spectroscopy at Okayama Astrophysical Observatory \\
2007 & 59, SP1 & 30 & 30 & 4.69 & First Results from Suzaku \\
2007 & 59, SP2 & 19 & 15 & 1.39 & AKARI Initial Results \\
2007 & 59, SP3 & 43 & 43 & 1.21 & Initial Results from Hinode \\
2008 & 60, SP1 & 35 & 35 & 1.13 & Recent Results from Suzaku \\
2009 & 61, SP1 & 37 & 37 & 0.97 & New Results from Suzaku \\
2009 & 61, SP2 & 1 & 1 & 1.94 & Survey of Period Variations of Superhumps in SU UMa-Type Dwarf Novae \\
2015 & 67, 3 & 5 & 5 & 0.98 & Subaru Never Sleeps \\
2015 & 67, 5 & 8 & 8 & 0.97 & Subaru Never Sleeps (continued from Vol. 67  3) \\
2018 & 70, SP1 & 40 & 40 & 4.49 & Subaru Hyper Suprime-Cam Survey \\
2018 & 70, SP2 & 20 & 19 & 1.40 & Star Formation Triggered by Cloud--Cloud Collision \\
2021 & 73, 1 & 4 & 4 & 1.17 & OISTER I: The Optical and Infrared Synergetic Telescopes for Education and Research \\
2021 & 73, SP1 & 21 & 21 & 1.28 & Star Formation Triggered by Cloud--Cloud Collision II \\
\hline
\end{tabular}
\par}
\begin{tabnote}
\footnotemark[1] Number of papers published in the Special Issues/Features in \pasj\ retrieved from the PASJ website https://www.asj.or.jp/pasj/rev-sp/rev-sp.html.  \\ 
\footnotemark[2] Number of papers indexed as ``Article'' publications in Scopus/SciVal.  
\end{tabnote}
\end{table*}

The indicators in Section~\ref{sec:3} position \pasj\ as a steady, mid-volume venue with comparatively low international co-authorship and a modest citation standing in the global landscape. CPP for older publication years lies below that of \apj, \mnras, and \aap. FWCI is typically below 1.0, with a few episodic peaks. Cumulative citation distributions show a steeper rise at low citation counts and a thinner upper tail than those of the largest general journals.
Within this profile, \pasj\ continues to serve as a platform for astronomy led by Japan and partners, including observational results, instrumentation, surveys, and methods. The field's 10--15-year maturation window suggests that durable, reusable research assets can accumulate influence over extended horizons. 

One opportunity is the deliberate publication of legacy articles that are widely cited and catalyze subsequent research (e.g., data products and value-added catalogs). In some years, \apjs\ has an annual output comparable to \pasj\ but places a relatively greater emphasis on legacy content, which is associated with very high citation impact. 
Table~\ref{tab:pasj_special_issues} compiles \pasj\ special issues and features from 1996 to 2023 whose average FWCI is $\gtrsim 1.0$. 
To identify papers published in the special issues and features in \pasj\, we used the official \pasj\ website\footnote{``PASJ Invited Reviews and Special Issues/Features'' posted on the Astronomical Society of Japan's website $\langle$https://www.asj.or.jp/pasj/rev-sp/rev-sp.html$\rangle$.}.
From this list, we constructed the set of papers for each special issue or feature and computed the average FWCI using SciVal.
Themes centered on facilities and missions, particularly public data releases and early results, are prominent (e.g., ASCA (1996), Suzaku (2007), and the Subaru Telescope's Hyper Suprime-Cam Survey (2018)), with several issues achieving average FWCI well above the world baseline (e.g., 3--5).
Given \pasj's comparatively small annual volume, such collections can shift year-level metrics and enhance the journal's visibility.

A second opportunity is to raise the share of international co-authorship incrementally. Prior work consistently finds that international co-authorship is associated with higher (field-normalized) citation impact, although effect sizes vary and part of the advantage reflects team size, selection, and visibility, and that returns may plateau at very high partner counts \citep[e.g.,][]{Glanzel2001,ChinchillaRodriguez2019}. These findings suggest that increasing \pasj's international co-authorship share could be beneficial, while acknowledging that the observed link with citation impact reflects association rather than causation. With increased participation in international facilities (e.g., through the Atacama Large Millimeter/submillimeter Array and other large collaborations), further gains in visibility are plausible.

\section{Conclusion}\label{sec:5}

We compared eight astronomy journals in 1996--2024 using indicators from Scopus via SciVal and analyzed time series, citation distributions, and citation age profiles. The age profiles reveal a long knowledge-use cycle (rise within $\sim$2--4 years, near saturation by $\sim$10--12 years), cautioning against overreliance on short-window metrics. Within this landscape, \pasj\ generally lies below the world baseline yet shows episodic above-baseline impact via facility- or mission-driven special issues and features. Practical levers for \pasj\ include periodic, facility-anchored issues and strengthened international collaboration. Future work will cross-validate with other databases, extend to systematic article-level analyses, and compare age-dependent citation dynamics across fields.

\begin{ack}
The author thanks Drs.\ Kazuyuki Suzuki, Marc Hansen, and Michiaki Yumoto for their generous guidance on bibliometric methods, which greatly helped shape the foundation of this study. 
The author dedicates this paper to his son Kimihiro and daughter Shiori, who kindly shared a wonderful time no metric could capture during their birthday celebrations in the days leading up to the initial submission.
\end{ack}


\begin{thebibliography}{}

\bibitem[Abazajian et al.(2009)]{Abazajian2009} Abazajian, K.~N. et al.\ 2009, \apjs, 182, 543

\bibitem[Abt(1981)]{Abt1981} Abt, H.~A.\ 1981, \pasp, 93, 207

\bibitem[Abt(1996)]{Abt1996} Abt, H.~A.\ 1996, \pasp, 108, 1059

\bibitem[Abt(2000)]{Abt2000} Abt, H.~A.\ 2000, Scientometrics, 48, 65

\bibitem[Bornmann(2012)]{Bornmann2012}
Bornmann, L.\ 2012, EMBO Rep., 13, 673

\bibitem[Burton \& Kebler(1960)]{BurtonKebler1960} 
Burton, R.~E. \& Kebler, R.~W.\ 1960, American Documentation, 11, 18

\bibitem[Chinchilla-Rodríguez et al.(2019)]{ChinchillaRodriguez2019} Chinchilla-Rodr\'{i}guez, Z., Sugimoto, C.~R. \& Larivi\`{e}re, V.\ 2019, PLoS ONE 14, 6, e0218309.

\bibitem[Fujiwara(2025)]{Fujiwara2025} Fujiwara, H.\ 2025, \pasj, 77, XXX (in press)

\bibitem[Gl\"{a}nzel \& Schubert(2001)]{Glanzel2001} Gl\"{a}nzel, W., \& Schubert, A.\ 2001, Scientometrics 50, 199

\bibitem[Moed(2010)]{Moed2010}
Moed, H.~F.\ 2010, Journal of Informetrics, 4, 265

\bibitem[Mongeon \& Paul-Hus(2016)]{Mongeon2016} Mongeon, P. \& Paul-Hus, A.\ 2016, Scientometrics, 106, 213

\bibitem[Peterson(1988)]{Peterson1988}
Peterson, C.~J.\ 1988, \pasp, 100, 106

\bibitem[Spergel et al.(2003)]{Spergel2003} Spergel, D.~N., Verde, L., Peiris, H.~V., et al.\ 2003, \apjs, 148, 175

\bibitem[Waltman(2016)]{Waltman2016}
Waltman, L.\ 2016, Journal of Informetrics, 10, 365


\end{thebibliography}
\end{document}